\documentclass[aps,pra,showpacs,preprint,floatfix,12pt]{revtex4-1}

\usepackage{hyperref}
\hypersetup{colorlinks=true,linkcolor=blue,citecolor=blue,urlcolor=blue}

\usepackage{mathrsfs}
\usepackage{graphicx}
\usepackage{epstopdf}
\usepackage{amsmath}
\usepackage{float}
\usepackage{latexsym}
\usepackage{amsfonts}
\usepackage{amssymb}
\usepackage{array}
\usepackage{epsfig}
\usepackage{subfigure}
\usepackage{braket}

\newcommand{\beq}{\begin{equation}}
\newcommand{\eeq}{\end{equation}}
\newcommand{\ba}{\begin{array}}
\newcommand{\ea}{\end{array}}
\newcommand{\bea}{\begin{eqnarray}}
\newcommand{\eea}{\end{eqnarray}}

\newcommand{\elc}[6]{\text{\framebox{$#1 #2$}\hspace{-0.4pt}\framebox{$#3 #4$}\hspace{-0.4pt}\framebox{$#5 #6$}}}

\begin{document}

\title[]{Dynamics of atomic spin-orbit-state wave packets produced by
short-pulse laser photodetachment}
\author {S. M. K. Law}
\author{G. F. Gribakin}
\email[]{g.gribakin@qub.ac.uk}
\affiliation{Center for Theoretical Atomic, Molecular and Optical Physics,
School of Mathematics and Physics, Queen's University Belfast,
Belfast BT7 1NN, United Kingdom}

\begin{abstract}
We analyse the experiment by Hultgren \textit{et al.} [Phys. Rev. A
\textbf{87}, 031404 (2013)] on orbital alignment and quantum beats in
coherently excited atomic fine-structure manifolds produced by short-pulse
laser photodetachment of C$^-$, Si$^-$ and Ge$^-$ negative ions, and derive a
formula that describes the beats. Analysis of the experimental data enables us
to extract the non-coherent background contribution for each species, and
indicates the need for a full density matrix treatment of the problem.
\end{abstract}
\maketitle

\section{Introduction}

In this paper we analyse the dynamics of quantum wave packets produced
by coherent excitation of atomic fine-structure manifolds in laser
photodetachment experiments, and probed by multiphoton ionization. We show
that for atoms with $np^2~^3P$ ground state the corresponding signal has a
very specific shape, which we determine analytically and find to be in good
agreement with experiment.

The development of laser pulses of few-femtosecond duration allows one to
resolve the electron motion in valence shells of atoms and molecules in the
time domain (see, e.g.,
\cite{Young06,Rohringer09,Lysaght09,Argenti10,Goulielmakis10,Fleischer11}).
Recently 100~fs pulse pump-probe experiments  \cite{Hultgren13,Eklund13} were
carried out to investigate the dynamics induced by the spin-orbit
interaction in neutral atoms.
In these experiments C, Si and Ge atoms with outer $np^2$ configuration were
prepared in the $^3P$ ground state by photodetachment of the respective
half-filled valence shell negative ions ($np^3~^4S$) by a linearly polarized
pump pulse.
Upon interaction with the infrared laser pulse, the emission of $p$
electrons with orbital angular momentum projection $m=0$ is strongly favoured
in comparison to $m=\pm1$. This causes the formation of a state with an
electron density hole localized along the pump laser polarization axis and
constitutes the orbital alignment effect. Such a state is not an eigenstate
of the atomic Hamiltonian when the spin-orbit interaction is included,
but a superposition of the fine-structure levels $^3P_J$, which evolves in time
according to the energy splittings in this manifold. This means that even for
a light atom, such as C, the relativistic spin-orbit interaction is essential
in determining the electron dynamics following the pump pulse.
In Refs. \cite{Hultgren13,Eklund13} this effect was probed by applying a
time-delayed ionizing probe pulse and measuring the signal of ionized electrons
for parallel and perpendicular polarization of the pump and probe pulses, as a
function of the time delay.

The experimental findings demonstrated the dependence of the ionization yield
on the time-varying hole density and the presence of quantum beat oscillations
of the signal with the delay time for C and Si (with no distinct signal for
Ge). This showed that electron dynamics resulting from the spin-orbit
interaction could be observed for both lighter and heavier atoms. In a recent
paper \cite{Rey14} Rey and van der Hart used $R$-matrix theory with time
dependence (RMT) to model the experiment of Refs. \cite{Hultgren13,Eklund13}.
They calculated the electron spectra following ionization of carbon in the
initial orbitally-aligned states with magnetic quantum numbers $M_L=0$ and 1,
and observed significant differences between these two cases, matching the
experimental findings. They also considered the evolution of the
fine-structure-state wave packet with the pump-probe delay time and simulated
the experimental signal by integrating electron emission within the cone of
11.7 degrees around the polarization direction of the probe pulse with momenta
$p\geq 0.4$~a.u. The scaled normalized yield obtained in this way was found to
be in good agreement with the experimental data from Ref.~\cite{Eklund13}.

In the present work we show that the experimental results can be described in
a much simpler manner. We use the assumption (similar to that used in 
Ref.~\cite{Rey14} and key to the experimental method of
Refs. \cite{Hultgren13,Eklund13}) that removal of $m=0$ electrons dominates
both the photodetachment (pump) and subsequent photoionization (probe),
and consider the motion of the $np^2~^3P$ fine-structure wave packet in C, Si,
or Ge. This gives a simple analytical expression for the signal as a function
of the pump-probe time delay. In order to make comparisons with experimental
results of Ref. \cite{Hultgren13} we scale the signal to account for the
background counts that could be present under the experimental conditions.
This allows us to analyse the contribution of background to the observed signal
beats and to effectively describe the loss of coherence in the wave packets
for systems with short beat periods.

The structure of the paper is as follows. In Sec.~\ref{sec:th} we derive the
expression that describes the beats of the signal due to the time evolution of
the atomic-state wave packet. In Sec.~\ref{sec:results} we use our analytical
expression to model the experimental data and compare with the phenomenological
simulation used in Refs.~\cite{Hultgren13,Eklund13}. Section \ref{sec:conc}
provides brief conclusions. Note that we use atomic units throughout.

\section{Theory}\label{sec:th}

Before the arrival of the pump pulse the negative ions are in the $np^3~^4S$
ground state with the total orbital angular momentum $L=0$ and spin $S=3/2$, 
and their projections $M_L=0$ and $M_S=-3/2,\dots , 3/2$. The total angular
momentum and its projection are $J=S$ and $M=M_S$.

We assume that after an instantaneous photodetachment of an electron with
$m=0$, the atoms are produced in a $np^2~^3P$ state at zero time delay,
after which this state evolves according to the energy splittings of
the fine-structure manifold.
The initial state of the atom is described by its total orbital and spin
angular momentum quantum numbers $L$ and $S$ with projections $M_L$ and $M_S$,
respectively, which we denote by $\ket{L,M_L;S,M_S}$ (or a superposition
of such states, see below). In the $LS$-coupling scheme, the time evolution of
the initial atomic state $\ket{L,M_L;S,M_S}$ is given by
\begin{equation}\label{eq:JM}
\ket{\Psi (t)}=\sum_J e^{-iE_Jt} C_{LM_LSM_S}^{JM} \ket{J,M},
\end{equation}
where $\ket{J,M}$ is the fine-structure energy eigenstate with the
total angular momentum $J$, projection $M=M_L+M_S$, and energy $E_J$,
$C_{LM_LSM_S}^{JM}$ denotes a Clebsch-Gordan coefficient, and
$\ket{\Psi (0)}=\ket{L,M_L;S,M_S}$.

To find the occupancies of the electron orbitals with $m=0$ in
$\ket{\Psi(t)}$, which determine the ionization signal after the probe
pulse, we expand the fine-structure states $\ket{J,M}$ in the basis of $LS$
states $\ket{L,M_L;S,M_S}$ (see the Appendix for the explicit form of these in
terms of the single-particle states),
\begin{equation}\label{eq:evolve}
\ket{\Psi(t)}=\sum_J \sum_{M'_L,M'_S} e^{-iE_Jt} C_{LM_LSM_S}^{JM}
C_{LM'_LSM'_S}^{JM} \ket{L,M'_L;S,M'_S},
\end{equation}
where the second sum is over all $M'_L$ and $M'_S$ such that $M'_L+M'_S=M$.

The removal of an $m=0$ electron from the initial $M_S=3/2$ anion state
produces the atomic state $\ket{1,0;1,1}$. The subsequent time evolution of the
atomic wave packet is found by applying Eq. (\ref{eq:evolve}) and
evaluating the appropriate Clebsch-Gordan coefficients,
\begin{equation}\label{eq:evolve1}
\ket{\Psi _1(t)}=\frac{1}{2}(e^{-iE_2t}+e^{-iE_1t})\ket{1,0;1,1}
+\frac{1}{2}(e^{-iE_2t}-e^{-iE_1t})\ket{1,1;1,0}.
\end{equation}
The corresponding ionization signal after the probe pulse is proportional
to the probability of finding an $m=0$ electron in the state (\ref{eq:evolve1}),
\begin{align}\label{eq:Spar1}
S_\|^{(1)}&= \frac{1}{2}(1-\cos \omega _{21}t),\\ \label{eq:Sper1}
S_\perp ^{(1)}&= \frac{1}{4}(3+\cos \omega _{21}t),
\end{align}
for the parallel and perpendicular polarization of the probe, respectively.
Here $\omega _{JJ'}=E_J-E_{J'}$, and the explicit forms of the atomic states
given in the Appendix were used.

The removal of an $m=0$ electron from the initial $M_S=1/2$ anion state
produces a superposition of atomic states,
\begin{equation}\label{eq:12}
\frac{1}{\sqrt{3}}\ket{1,0;1,1}+\sqrt{\frac{2}{3}}\ket{1,0;1,0},
\end{equation}
(see the Appendix), whose time evolution is given by
\begin{align}\label{eq:evolve2}
\ket{\Psi_2(t)}&=\frac{1}{\sqrt{3}}\ket{\Psi _1(t)}+
\sqrt{\frac{2}{3}}\left( \frac{2}{3}e^{-iE_2t}+\frac{1}{3}e^{-iE_0t}\right)
\ket{1,0;1,0} \nonumber \\
&+\sqrt{\frac{2}{3}}\left( \frac{1}{3}e^{-iE_2t}-\frac{1}{3}e^{-iE_0t}\right)
(\ket{1,1;1,-1}+\ket{1,-1;1,1}).
\end{align}
This gives the probabilities of finding an $m=0$ electron at time $t$ as
\begin{align}\label{eq:Spar2}
S_\|^{(2)}&= \frac{1}{6}(1-\cos \omega _{21}t)
+\frac{8}{27}(1-\cos \omega _{20}t),
\\ \label{eq:Sper2}
S_\perp ^{(2)}&= \frac{1}{12}(7+\cos \omega _{21}t)
+\frac{1}{27}(5+4\cos \omega _{20}t).
\end{align}
Note that since the Clebsch-Gordan coefficient $C_{1010}^{10}$ is zero,
no interference is observed between the $J=0$ and $J=1$ sublevels. This is in
agreement with the experimental analysis \cite{Hultgren13,Eklund13}, which
allowed for the presence of the $\omega_{10}=E_1-E_0$ beat frequency, but
found its contribution statistically insignificant. Note also that the sum
$S_\|^{(i)}+2S_\perp ^{(i)}=2$, independently of time, which
could be expected since there are two orthogonal directions perpendicular to the
polarization of the pump pulse.

The total signals for the parallel and perpendicular polarizations
of the pump and probe pulses are proportional to $S_\|=S_\|^{(1)}+S_\|^{(2)}$
and $S_\perp = S_\perp ^{(1)}+S_\perp ^{(2)}$. (The initial anion states with
$M_S=-3/2$ and $-1/2$ give the identical contribution.) The normalized electron
yield measured in the experiment is
\begin{equation}\label{eq:signalformula}
S(t)=\frac{S_\perp-S_\|}{S_\perp+S_\|}.
\end{equation}
Using Eqs.~(\ref{eq:Spar1}), (\ref{eq:Sper1}), (\ref{eq:Spar2}), and
(\ref{eq:Sper2}), and allowing for a constant time shift $t_0$ related to the
uncertainty of the zero time delay \cite{Eklund13}, and for some signal
background that may contribute to $S_\|$ and $S_\perp $, we obtain
\begin{equation}\label{eq:signal}
S(t)=\frac{\dfrac{5}{9}+\cos[\omega_{21}(t-t_0)]+
\dfrac{4}{9}\cos[\omega_{20}(t-t_0)]+\Delta S_b}
{\dfrac{67}{27}-\dfrac{1}{3}\cos [\omega_{21}(t-t_0)]-
\dfrac{4}{27}\cos[\omega_{20}(t-t_0)]+S_b},
\end{equation}
where $S_b$ and $\Delta S_b$ are the sum and difference of the
background contributions for the parallel and perpendicular polarizations.
Such background can also account for reduced coherence of the wave packet
when the pump pulse duration becomes comparable to or greater than the beat
periods (see Sec.~\ref{sec:results}). Note that a better quantity than
that in Eq.~(\ref{eq:signalformula}) would probably be the ratio $(2S_\perp -S_\| )/(2S_\perp +S_\|)$, in which the denominator should be constant.

The contribution of the oscillating terms in the denominator of
Eq.~(\ref{eq:signal}) is relatively small, even in the absence of any background
$S_b$. This means that $S(t)$ is close to a simple linear combination of
a constant and two beat components with frequencies $\omega_{21}$ and
$\omega_{20}$. Equation (\ref{eq:signal}) shows that the relative
contribution of the beats with frequencies $\omega_{21}$ and $\omega_{20}$ is
fixed, and the beat period between the higher-lying levels
$J=1,\,2$ gives the dominant contribution. Note also that Eq.~(\ref{eq:signal})
with $\Delta S_b=S_b=0$ predicts a positive constant offset $\overline{S}=
15/67\sim 0.2$, which is qualitatively similar to the observations (see
Sec.~\ref{sec:results}).

In analysing the experimental data, the authors of
Refs.~\cite{Hultgren13,Eklund13} used the following phenomenological function
\begin{equation}\label{eq:phenomenological}
f(t)=c_0+\alpha_1 \cos[\omega_{21}(t-t_0)]+\alpha_2 \cos[\omega_{20}(t-t_0)]+
\alpha_3 \cos[\omega_{10}(t-t_0)],
\end{equation}
with five fitting parameters: a constant offset $c_0$, amplitudes
$\alpha_i$ ($i=1,\,2,\,3$) of all three possible beats with
frequencies $\omega_{21}$, $\omega_{20}$, and $\omega_{10}$,
and $t_0$.

The beat frequencies are determined by the corresponding energy splittings
\cite{NIST}: $\omega_{21}=5.086$, 27.510, and 160.65~ps$^{-1}$,
$\omega_{20}=8.175$, 42.028, and 265.56~ps$^{-1}$, and $\omega_{10}=3.089$,
14.524, and 104.95~ps$^{-1}$, for C, Si, and Ge, respectively. The corresponding
beat periods are $\tau_{JJ'}=2\pi /\omega_{JJ'}$.
In Sec. \ref{sec:results} we compare the results obtained
using our three-parameter fits (\ref{eq:signal}) with those of
Eq.~(\ref{eq:phenomenological}).

\section{Results and discussion}\label{sec:results}

\begin{figure}[ht]
\hspace*{-0.5in}
\mbox{\subfigure{\includegraphics[width=3in]{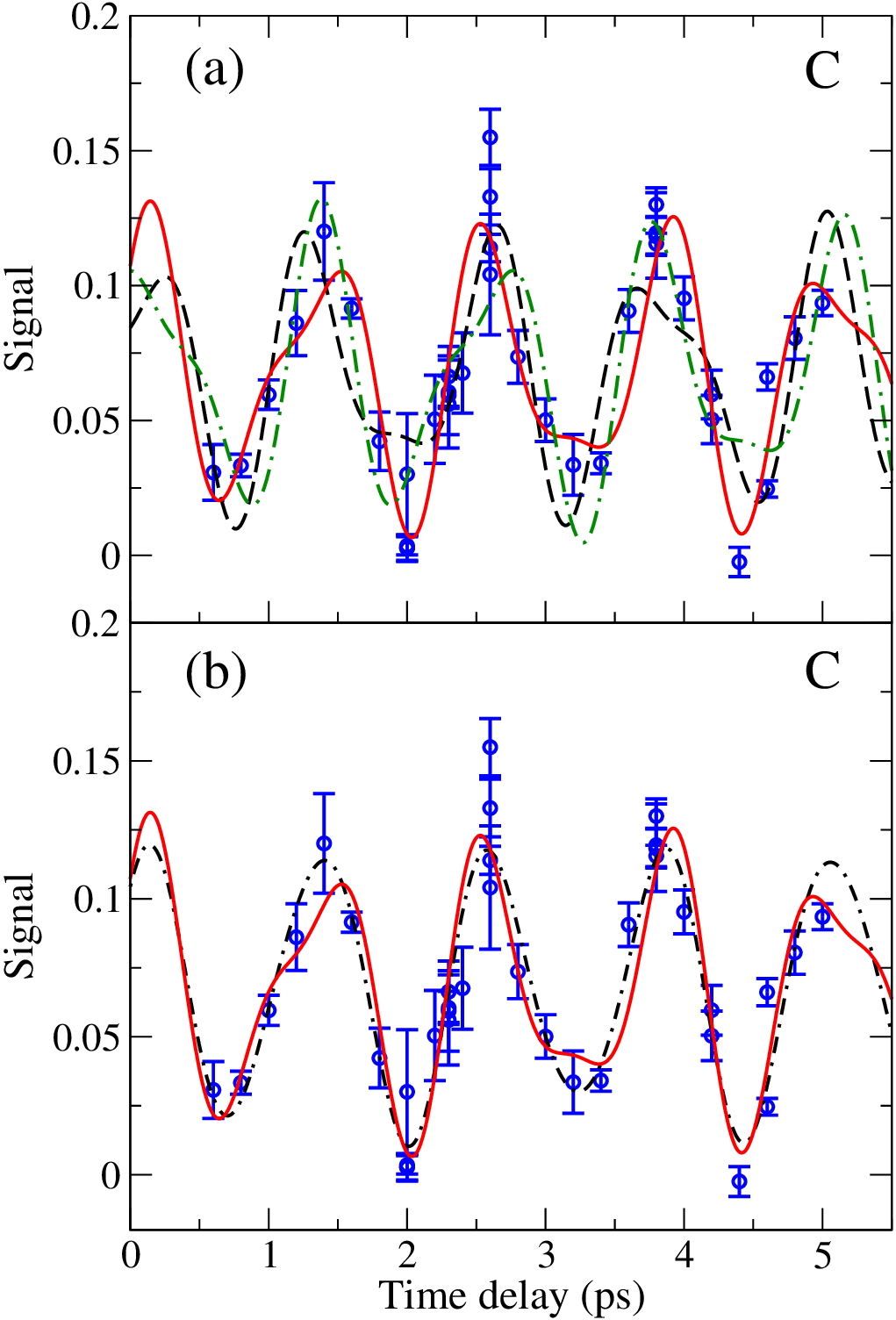}}\quad
\subfigure{\includegraphics[width=3in]{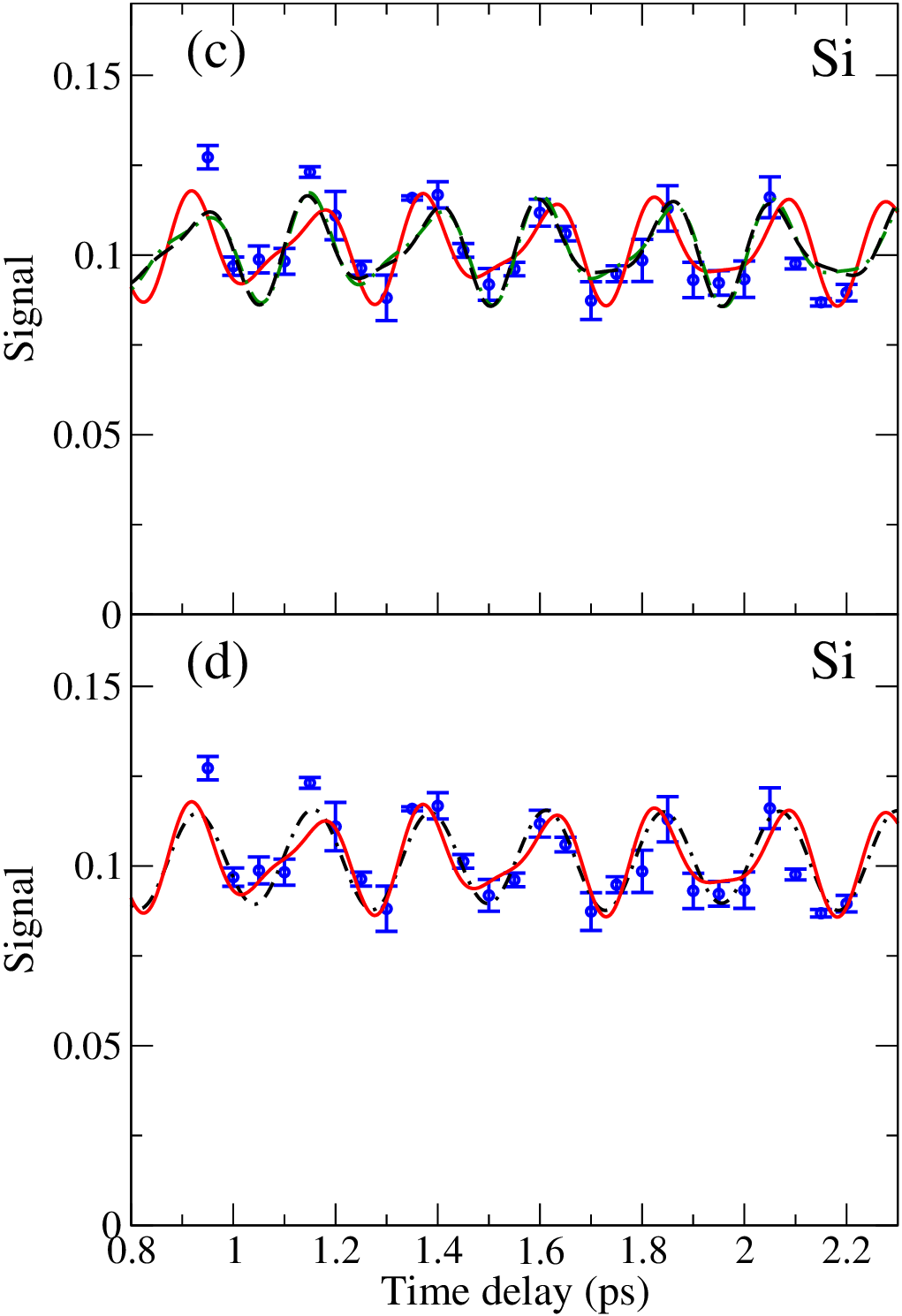}}}
\caption{(Color online) Normalized electron ionization yield as a function of
time delay for C [panels (a) and (b)] and Si [panels (c) and (d)]. The graphs
in (a) and (c) show several theoretical fits using Eq. (\ref{eq:signal}) with
different time shift parameter $t_0$ that best model
the experimental data \cite{Eklund13} plotted in blue circles in each panel.
The three fits for C in panel (a) correspond to $t_0=-1.127$~ps (black dashed
line), 0.144~ps (red solid line) and 1.381~ps (green dash-dotted line).
For Si in panel (c) the fits correspond to $t_0=-0.213$~ps (black dashed line),
0.013~ps (red solid line) and 0.244~ps (green dash-dotted line). Other
parameters are listed in Table~\ref{tab:background}. Panels (b) and (d)
compare the best fit from (a) and (c) for C and Si, respectively (red solid
line) with the fit obtained by using Eq.~(\ref{eq:phenomenological})
\cite{Hultgren13} (black dash-dotted line).}
\label{fig:csi}
\end{figure}

\begin{figure}[ht]
\includegraphics[width=3.8in]{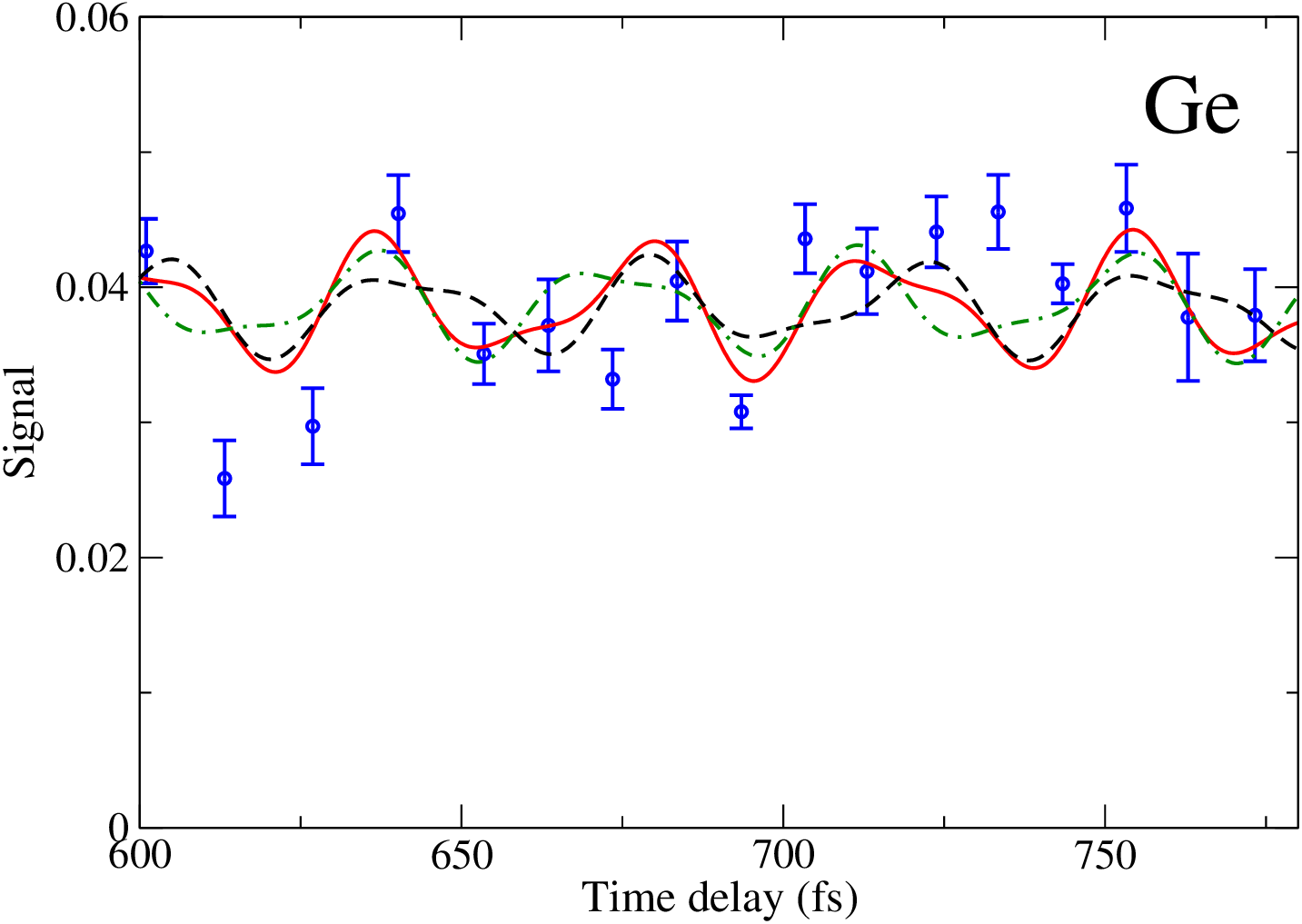}
\caption{(Color online) Normalized electron yield as a function of
time delay for Ge. Experimental data \cite{Eklund13} (blue circles); fits
using Eq. (\ref{eq:signal}) correspond to $t_0=-27.10$~fs (solid red line),
15.66~fs (dashed black line), and 47.98~fs (dash-dotted green line); other
parameters are in Table~\ref{tab:background}.}
\label{fig:ge}
\end{figure}

Figure \ref{fig:csi} displays the results for the normalized yield $S(t)$,
Eq.~(\ref{eq:signalformula}), as a function of time delay for carbon
[panels (a) and (b)] and silicon [panels (c) and (d)]. The experimental
results from Ref.~\cite{Eklund13}, obtained from momentum-resolved images for
high-energy ($p\geq 0.4$~a.u.) ionized electrons, are shown by blue circles in
each panel. They are compared with (i) our
analytical formula (\ref{eq:signal}) fitted using $t_0$, $S_b$, and
$\Delta S_b$ as free parameters, and (ii) the phenomenological
five-parameter fit Eq.~(\ref{eq:phenomenological}) used previously in Refs.
\cite{Hultgren13,Eklund13}. By varying the range of the time shift parameter
$t_0$ in Eq.~(\ref{eq:signal}), several locally optimal fits may be achieved,
the one with the smallest absolute value of $t_0$ being the overall best
(shown by the solid red line). Figure~\ref{fig:ge} shows three fits for Ge,
using only Eq.~(\ref{eq:signal}) and plotted similarly to Fig.~\ref{fig:csi}
(a) and (c). A full list of fitted parameter values is in
Table~\ref{tab:background}.

\begin{table}[ht!]
\caption{Values of parameters $t_0$, $\Delta S_b$, and $S_b$ used in
Eq.~(\ref{eq:signal}) to fit the experimental data from Ref.~\cite{Eklund13} in
Figs. \ref{fig:csi} and \ref{fig:ge}. The $\chi^2$ error representing the
quality of the fit for each set of parameters is also shown. The choice of
parameters that gives the best fit for each atom is shown in bold.}
\label{tab:background}
\begin{ruledtabular}
\begin{tabular}{ccccc}
Atom & & Best-fit parameters & & Error \\
\cline{2-4}
 & $\Delta S_b$ & $S_b$ & $t_0$ (ps) & $\chi^2$ \\
\hline
C & 1.11 & 22.30 & $-1.127$ & $1.98\times 10^{-2}$ \\
 & ${\bf 1.02}$ & ${\bf 20.96}$ & ${\bf 0.144}$ & ${\bf 8.89 \times 10^{-3}}$ \\
 & 0.97 & 20.45 & 1.381 & $1.67\times 10^{-2}$ \\
\hline
Si & 8.88 & 90.38 & $-0.213$ & $1.39\times 10^{-3}$\\
 & ${\bf 8.77}$ & ${\bf 88.88}$ & ${\bf 0.013}$ & ${\bf 1.35\times 10^{-3}}$\\
 & 8.85 & 89.96 & 0.244 & $1.38\times 10^{-3}$ \\
\hline
Ge & ${\bf 9.40}$ & ${\bf 255.18}$ & ${\bf -0.027}$ &
${\bf 4.33\times 10^{-4}}$ \\
 & 13.51 & 362.73 & 0.016 & $5.13\times 10^{-4}$ \\
 & 12.10 & 324.13 & 0.048 & $4.79\times 10^{-4}$ \\
\end{tabular}
\end{ruledtabular}
\end{table}

Simulation of the experimental data by means of Eq.~(\ref{eq:signal}) 
for C and Si clearly shows the temporal oscillations originating from quantum
beat interference between the coherently populated $J$ sublevels of the $^3P$
ground state. These oscillations are faster for heavier atoms, as observed in
the experiment, which is directly linked to the decrease in the spin-orbit
periods
$\tau_{JJ'}$ for larger fine-structure splitting energies $\omega_{JJ'}$. As
predicted by Eq.~(\ref{eq:signal}), the observed beat periods are dominated by
the $J=1,\,2$ sublevel contributions with $\tau_{21}=1.24$~ps,
$\tau_{21}=0.23$~ps and $\tau_{21}=39.11$~fs for C, Si, and Ge, respectively. 

By introducing the appropriate background parameters in the fits, good
agreement with the experimental data is observed in Fig. \ref{fig:csi} for C
and Si. The best fits [shown by solid red curves in Fig. 1 (b) and (d)]
correspond to the smallest absolute values of $t_0$ ($\lesssim 100$~fs), other
$t_0$ values
differing from it by $\sim \tau_{21}$. For both atoms, the beat pattern is
dominated by the $\tau _{21}$ period component, with the $\tau _{20}$
beat component producing a characteristic ``knee'' visible at even
half-periods. A similar pattern was observed in numerical simulations for C by
Rey and van der Hart \cite{Rey14}, but it is totally absent from the fit with
the function $f(t)$,
Eq.~(\ref{eq:phenomenological}), used in the experimental papers
\cite{Hultgren13,Eklund13} [black dash-dotted lines in Fig. 1 (b) and (d)].

For Ge, however, the oscillatory behaviour predicted by Eq.~(\ref{eq:signal})
does not provide a good description of the experimental data for any choice of
parameters (see Fig.~\ref{fig:ge}). A related feature of the data is that the
scale of the oscillations becomes very small in Ge compared with C and Si. This
can be seen from the fitted values of $S_b$ and $\Delta S_b$
in Table~\ref{tab:background}. From Eq.~(\ref{eq:signal}), the time-independent
part of the asymmetry is determined by the ratio
$\overline{S}=(5/9+\Delta S_b)/(67/27+S_b)\sim \Delta S_b/S_b$, while the
amplitude of the
beats is $\sim 1/S_b$. The data in Figs.~\ref{fig:csi} and \ref{fig:ge}, and in
Table~\ref{tab:background} show that $\Delta S_b/S_b\sim 0.05$--0.1 for all
three
species, while the amplitude of the beats decreases from 0.05 for C, to 0.01
for Si, and $4\times 10^{-3}$ for Ge. The latter value is close to the size of
error bars in the experimental data for Ge.

This behaviour is related to the effect of the pulse duration in comparison
with the beat periods. In the derivation of Eq.~(\ref{eq:signal}), the
removal of $m=0$ electrons was assumed to be instantaneous, leading to fully
coherent (pure) time-dependent states with wave functions (\ref{eq:evolve1})
and (\ref{eq:evolve2}). 
In the experiment \cite{Hultgren13,Eklund13} the duration of the pump and probe
pulses was 100~fs, which is much shorter the main beat
period for C and shorter than that for Si, but is 2.5 times greater than
$\tau_{21}$ for Ge. As a result, the degree of coherence in the spin-orbit
wave packet is largest for C, but becomes progressively smaller in Si and Ge.
This results in the reduction of the coherent (oscillatory) part of the signal,
with the atomic states produced by the pump becoming closer to a classical
ensemble rather than a quantum superposition.

The spin-orbit wave packet in Ge (and to a lesser extent, in Si) is also
affected by strong dependence of the multiphoton detachment rates on
the threshold energy. This leads to a greater suppression of the detachment
probability for higher-lying final atomic states with $J=1$ and 2, compared with
that for the $J=0$ ground state. The magnitudes of the lowest ($J=0$) and
highest ($J=2$)
thresholds are 1.2621 and 1.2675~eV in C$^-$, 1.3895 and 1.4172~eV in Si$^-$,
and 1.2327 and 1.4075~eV in Ge$^-$. Using the method of Ref.~\cite{Gribakin97},
we estimate that for a laser pulse with wavelength $\lambda =2055$~nm (as in
Refs.~\cite{Hultgren13,Eklund13}) and intensity $I=2\times 10^{12}$~W/cm$^2$
(for which the total detachment probability over 100~fs is close to unity),
the increase in the threshold energy from the $J=0$ to $J=2$ state leads to
2\%, 7\%, and 40\% reduction of the detachment rate, for C, Si, and Ge,
respectively. As a result, the contribution of the $J=2$ state to the
wavepacket (\ref{eq:JM}), which is critical for the magnitude of the beats, 
can be reduced below the values predicted by the $LS$-coupling coefficients.

The above analysis makes it clear that a complete description of the beat
character and spin-orbit coherences of the atomic ensemble requires a density
matrix consideration of the problem \cite{Rohringer09,Goulielmakis10,Law16}.
Depending on the pump pulse length, strong-field detachment may not generally
produce perfectly coherent aligned states.
The elements of the density matrix in the $\ket{J,M}$ basis can be determined
by calculating the detachment amplitudes for a variety of pulse lengths using
existing theory of strong-field photodetachment (e.g., Keldysh-type
theory \cite{Keldysh64,Gribakin97,Shearer11,Shearer13,Korneev12}). In this
approach the diagonal elements will represent populations of different atomic
fine-structure levels for a given $M$, and the magnitude of the off-diagonal
elements will describe coherences between the $J$ states. The degree
of coherence is then described by the ratio of the off-diagonal elements to
the geometric mean of the corresponding diagonal elements. Calculations for
the halogen negative ions, whose detachment leads to two fine-structure 
atomic states, show that the degree of coherence is a function of the
ratio $\tau _p/\tau _{JJ'}$, where $\tau_p$ is the laser pulse length
\cite{Law16}. For $\tau _p/\tau _{JJ'}\ll 1$ the degree of coherence is
close to unity, but it drops quickly for $\tau _p/\tau _{JJ'}\sim1$ and
reaches few-percent values for $\tau _p\approx 2\tau _{JJ'}$, which is similar
to the situation in Ge.

\section{Conclusions}\label{sec:conc}

We have investigated the evolution of the ground-state spin-orbit wave packets
in carbon, silicon and germanium atoms produced by detachment of $m=0$ electrons
from half-filled valence $np^3$ negative ions. A simple analytical formula that
describes the time-changing alignment of electron orbitals, as probed in the
pump-probe experiment, has been derived and applied to the analysis of
experimental data \cite{Hultgren13,Eklund13}. For C and Si the theory
provides a good description of temporal beat oscillations which demonstrate the
existence of a coherent superposition of the fine-structure sublevels
of the atomic triplet state.
The sharp suppression of the coherence degree observed experimentally for Ge
demonstrates that the assumption of an instantaneous pulse is insufficient for
atoms with shorter beat periods (in the femtosecond range). This calls for a
full density-matrix consideration of the problem that would provide a
complete description of partially coherent dynamics occurring in spin-orbit
manifolds of general atoms with $l\geq 1$ valence electron orbitals.

\section{Acknowledgments}
The work of S.M.K.L. has been supported by the Department for Employment and
Learning, Northern Ireland. We thank I. Kiyan and M. Eklund for providing
experimental data and for useful discussions. 

\appendix*

\section{Valence-electron states of anions and atoms of C, Si, and Ge}

The possible initial states of the $np^3~^4S$ negative ion are
\begin{align}
\ket{0,0;3/2,3/2}&=\elc{\uparrow}{~}{\uparrow}{~}{\uparrow}{~}~,\\
\ket{0,0;3/2,1/2}&=\frac{1}{\sqrt{3}}\left(~
\elc{\uparrow}{~}{\uparrow}{~}{~}{\downarrow}+
\elc{\uparrow}{~}{~}{\downarrow}{\uparrow}{~}+
\elc{~}{\downarrow}{\uparrow}{~}{\uparrow}{~}~\right)~,
\end{align}
where each of the boxes represents a state of three electrons in the
$np$ orbital, with magnetic quantum numbers $m=-1$, 0, and 1, and up
($\uparrow $) or down ($\downarrow$) spins. The states with $M_S=-1/2$ and
$-3/2$ are similar and, owing to the symmetry with respect to reflection in
the $x$-$y$ plane, they need not be considered.

The two-electron $np^2~^3P$ states of the neutral atom that can be formed by
removal of an $m=0$ electron from the above states are
\begin{align}\label{eq:1011}
\ket{1,0;1,1}&=\elc{\uparrow}{~}{\vphantom{\uparrow}~~\,}{~}{\uparrow}{~}~,\\
\label{eq:1010}
\ket{1,0;1,0}&=\frac{1}{\sqrt{2}}\left(~
\elc{\uparrow}{~}{\vphantom{\uparrow}~~\,}{~}{~}{\downarrow}+
\elc{~}{\downarrow}{\vphantom{\uparrow}~~\,}{~}{\uparrow}{~}~\right)~,
\end{align}
or their superposition.
Other atomic states that appear in the $LS$-expansion of the fine-structure
levels $\ket{J,M}$ linked to the states (\ref{eq:1011}) and (\ref{eq:1010}),
are
\begin{align}\label{eq:1110}
\ket{1,1;1,0}&=\frac{1}{\sqrt{2}}\left(~
\elc{\vphantom{\uparrow}~~\,}{~}{\uparrow}{~}{~}{\downarrow}+
\elc{\vphantom{\uparrow}~~\,}{~}{\downarrow}{~}{\uparrow}{~}~\right)~,\\
\label{eq:1-111}
\ket{1,-1;1,1}&=\elc{\uparrow }{~}{\uparrow }{~}{\vphantom{\uparrow}~~\,}{~}~,\\
\label{eq:111-1}
\ket{1,1;1,-1}&=\elc{\vphantom{\uparrow}~~\,}{~}{\downarrow }{~}{\downarrow }{~}~.
\end{align}
Using states (\ref{eq:1011})--(\ref{eq:111-1}) it is straightforward to work
out the relative probabilities of removing $m=0$ electron by the ionizing probe
pulse with polarization parallel to the pump pulse.

For the perpendicular probe
polarization, one needs to expand the angular parts of the $np$ electron
wave functions $Y_{1m}(\theta ,\phi )$ in terms of the spherical functions in a
coordinate frame with the $z$ axis perpendicular to the original $z$ axis
\cite{Varsh}:
\begin{align}\label{eq:Y11}
Y_{11}(\theta ,\phi )&=\frac{1}{2}Y_{11}(\tilde \theta ,\tilde \phi )
-\frac{1}{\sqrt{2}}Y_{10}(\tilde \theta ,\tilde \phi )
+\frac{1}{2}Y_{1-1}(\tilde \theta ,\tilde \phi ),\\ \label{eq:Y10}
Y_{10}(\theta ,\phi )&=\frac{1}{\sqrt{2}}Y_{11}(\tilde \theta ,\tilde \phi )
-\frac{1}{\sqrt{2}}Y_{1-1}(\tilde \theta ,\tilde \phi ),\\ \label{eq:Y1-1}
Y_{1-1}(\theta ,\phi )&=\frac{1}{2}Y_{11}(\tilde \theta ,\tilde \phi )
+\frac{1}{\sqrt{2}}Y_{10}(\tilde \theta ,\tilde \phi )
+\frac{1}{2}Y_{1-1}(\tilde \theta ,\tilde \phi ).
\end{align}
Here $\tilde \theta $ and $\tilde \phi $ are the polar angles of the
new coordinate frame, obtained by rotation through 90 degrees about the
original $y$ axis. These formulae show that for the states 
(\ref{eq:1011}) and (\ref{eq:1010}) with $M_L=0$,
the average number of $m=0$ electrons detected in the perpendicular direction
is unity, while for the states (\ref{eq:1110})-(\ref{eq:111-1}) with
$M_L=\pm 1$ this number is 0.5. Alternatively, one can expand the
fine-structure states $\ket{J,M}$ in Eq.~(\ref{eq:JM}) in the frame with the
perpendicular $z$ axis using equations similar to
(\ref{eq:Y11})--(\ref{eq:Y1-1}), and analyse the time
evolution in it.

\end{document}